\documentclass[prl,showpacs,showkeys,nofootinbib,twocolumn,floatfix]{revtex4}

\usepackage{graphicx}  %
\usepackage{amsmath}
\usepackage{bm}  %

\newcommand{\bea}{\begin{eqnarray}}
\newcommand{\eea}{\end{eqnarray}}
\newcommand{\beq}{\begin{equation}}
\newcommand{\eeq}{\end{equation}}
\newcommand{\bqa}{\begin{eqnarray}}
\newcommand{\eqa}{\end{eqnarray}}

\def\mqo2{{\!\!\!}}

\usepackage{relsize}
\def\babar{\mbox{\slshape B\kern-0.1em{\smaller A}\kern-0.1em
    B\kern-0.1em{\smaller A\kern-0.2em R}}}

\begin{document}

\title{
Selection Rules for \\ 
Hadronic Transitions of $\bm{XYZ}$ Mesons}
\author{Eric Braaten}
\affiliation{Department of Physics,
         The Ohio State University, Columbus, OH\ 43210, USA\\}
\author{Christian Langmack}
\affiliation{Department of Physics,
         The Ohio State University, Columbus, OH\ 43210, USA\\}
\author{D.~Hudson Smith}
\affiliation{Department of Physics,
         The Ohio State University, Columbus, OH\ 43210, USA\\}
\date{\today}

\begin{abstract}
Many of the $XYZ$ mesons discovered in the last decade can be 
identified as bound states of a heavy quark and antiquark
in Born-Oppenheimer (B-O) potentials defined by the energy of 
gluon and light-quark fields in the presence of static color sources. 
The mesons include quarkonium hybrids, which are bound states in 
excited flavor-singlet B-O potentials, and quarkonium tetraquarks, 
which are bound states in flavor-nonsinglet B-O potentials. 
The deepest hybrid potentials are known from  lattice QCD calculations. 
The deepest  tetraquark potentials can be inferred from
lattice QCD calculations of static adjoint mesons.
Selection rules for hadronic transitions are derived and used to identify $XYZ$ mesons
that are candidates for ground-state energy levels in the B-O potentials
for charmonium hybrids and tetraquarks.
\end{abstract}

\smallskip
\pacs{14.40.Pq,14.40.Rt,31.30.-i,13.25.Gv}
\keywords{
Quarkonium, hybrid mesons, tetraquark mesons, exotic mesons,
Born-Oppenheimer approximation, hadronic transitions, selections rules. }
\maketitle

The $XYZ$ mesons are unexpected mesons 
discovered during the last decade that contain 
a heavy quark-antiquark pair and are above the open-heavy-flavor threshold.  
Recent discoveries of charged $XYZ$ mesons in both the 
$b \bar b$ sector and the $c \bar c$ sector
have established unambiguously the existence of tetraquark mesons
that contain two quarks and two antiquarks.
In 2011, the Belle Collaboration discovered $Z_b^+(10610)$ and $Z_b^+(10650)$, 
which are revealed by their decays into $\Upsilon\, \pi^+$
to be tetraquark mesons with constituents $b\bar b u \bar d$  \cite{Belle:2011aa}.
In 2013, the BESIII Collaboration discovered the $Z_c^+(3900)$, 
which is revealed by its decay into $J/\psi\, \pi^+$ to be a tetraquark meson 
with constituents $c\bar c u \bar d$ \cite{Ablikim:2013mio}.
An updated list of the $XYZ$ mesons as of August 2013 was given in 
Ref.~\cite{Bodwin:2013nua}.
The list included 15 neutral and 4 charged $c \bar c$ mesons.
The BESIII collaboration has recently observed an additional neutral state $Y(4220)$
 \cite{Yuan:2013ffw} and additional charged states
$Z_c^+(3885)$  \cite{Ablikim:2013xfr} and $Z_c^+(4020)$  \cite{Ablikim:2013wzq}.
The list in Ref.~\cite{Bodwin:2013nua} 
also included 1 neutral and 2 charged $b \bar b$ mesons.

A full decade has elapsed since the discovery of the first $XYZ$ meson,
the $X(3872)$ \cite{Choi:2003ue},
but no compelling explanation for the pattern of $XYZ$ mesons
has emerged.
In simple constituent models, an $XYZ$ meson 
consists of a heavy quark ($Q$) and antiquark ($\bar Q$)
and possibly additional constituents that could be gluons ($g$)
or light quarks ($q$) and light antiquarks ($\bar q$).
The models that have been proposed can be 
classified according to how the constituents are clustered
within the meson.  They include
(1) {\it conventional quarkonium}: $(Q \bar Q)_1$,
(2) {\it quarkonium hybrid meson}:  $(Q \bar Q)_8 + g$,
(3) {\it compact tetraquark}:  $(Q \bar Q q \bar q)_1$ \cite{Vijande:2007rf},
(4) {\it meson molecule}:  $(Q \bar q)_1 + (\bar Q q)_1$  \cite{Tornqvist:1993ng},
(5) {\it diquark-onium}:  $(Q q)_{\bar 3} + (\bar Q \bar q)_3$  \cite{Drenska:2010kg}, 
(6) {\it hadro-quarkonium}:  $(Q \bar Q)_1 + (q \bar q)_1$  \cite{Dubynskiy:2008mq},
and (7) {\it quarkonium adjoint meson}:  
$(Q \bar Q)_8 + (q \bar q)_8$   \cite{Braaten:2013boa}.
The subscripts indicate the color charge of the clusters within the meson.
All of these are possible models for neutral $XYZ$ mesons.
The last five are possible models for charged $XYZ$ mesons.
It would be desirable to have a theoretical framework firmly based on QCD that
describes all the $XYZ$ mesons,
including their masses, widths, quantum numbers, and decay modes.
The Born-Oppenheimer (B-O) approximation may provide such a  theoretical framework.

The B-O approximation is used in atomic 
and molecular physics to understand the binding of atoms into molecules. 
It exploits the large ratio of the time scale for the motion 
of the atomic nuclei to that for the electrons, which is a consequence 
of the large ratio of the nuclear and electron masses. 
The electrons respond almost instantaneously to the motion of the nuclei,
which can be described by the Schroedinger equation in 
a B-O potential defined by the energy of the electrons 
in the presence of static electric charges. 
The B-O approximation for $Q \bar Q$ mesons in QCD
was developed by Juge, Kuti, and Morningstar \cite{Juge:1999ie}. 
It exploits the large ratio of the time scale for the motion 
of the $Q$ and $\bar Q$ to that for the evolution of gluon fields, 
which is a consequence of the large ratio of the heavy-quark mass 
to the nonperturbative momentum scale $\Lambda_{\rm QCD}$.
The gluon field responds 
almost instantaneously to the motion of the $Q \bar Q$ pair,
which can be described by the Schroedinger equation in 
a B-O potential defined by the energy of the gluon field 
in the presence of static color sources.
Conventional quarkonia are energy levels of a $Q \bar Q$ pair 
in the ground-state B-O potential.
The energy levels in the excited-state B-O potentials are quarkonium hybrids.  
Juge, Kuti, and Morningstar calculated many of the B-O potentials 
using quenched lattice QCD \cite{Juge:1999ie}. 
They calculated the spectra of charmonium hybrids and bottomonium hybrids
by solving the Schroedinger equation in the B-O potentials.
They also calculated some of the bottomonium hybrid energies 
using lattice nonrelativistic QCD (NRQCD).  The quantitative agreement between
the predictions of the B-O approximation
and lattice NRQCD provided convincing evidence for the existence of
quarkonium hybrids in the hadron spectrum of QCD.

It is also possible to define flavor-nonsinglet B-O potentials by the energies 
of stationary configurations of light-quark and gluon fields with nonsinglet quark flavors  
in the presence of static color sources \cite{Braaten:2013boa}.
The energy levels of a $Q \bar Q$ pair in such a potential
are quarkonium tetraquarks.
The component of the wavefunction in which the separation of the $Q \bar Q$ pair
is much smaller than the spatial extent of the light-quark and gluon fields
resembles a quarkonium adjoint meson ($(Q \bar Q)_8 + (q \bar q)_8$).
Components in which the $Q$ and $\bar Q$ are well separated may resemble
a meson molecule ($(Q \bar q)_1 + (\bar Q q)_1$)  
or diquark-onium ($(Q q)_{\bar 3} + (\bar Q \bar q)_3$).

In this paper, we apply the B-O approximation for quarkonium 
hybrids and tetraquarks to the $c \bar c$ $XYZ$ mesons.
The deepest flavor-singlet B-O potentials have been determined using lattice QCD calculations.
We infer the deepest flavor-nonsinglet B-O potentials from lattice QCD calculations
of the static adjoint meson spectrum.
We derive selection rules for hadronic transitions
and use them to identify $XYZ$ mesons that are candidates 
for ground-state energy levels of charmonium hybrids and tetraquarks.

The B-O potentials for $Q \bar Q$  mesons can be labelled 
by quantum numbers for the gluon and light-quark fields that are conserved  
in the presence of static $Q$ and $ \bar Q$ sources separated by a vector
$\bm{r}$ \cite{Juge:1999ie}:
(1) the eigenvalue $\lambda = 0, \pm 1, \pm 2, \ldots$ of
$\hat{\bm{r}}\cdot \bm{J}_{\rm light}$, where $\bm{J}_{\rm light}$ 
is the total angular momentum vector for the light fields,  
(2) the eigenvalue $\eta = \pm 1$ of 
$(CP)_{\rm light}$, which is the product of the charge-conjugation 
operator and the parity operator that spacially inverts the light fields through 
the midpoint between the $Q$ and $\bar Q$ sources, 
(3) for the  case $\lambda = 0$, the eigenvalue $\epsilon = \pm 1$
of a reflection operator $R_{\rm light}$ that reflects the light fields
through  a plane containing the $Q$ and $\bar Q$ sources,
(4) quark flavors, which can be flavor-singlet or $q_1 \bar q_2$, 
where $q_1,q_2=u,d,s$. 
For $q_1,q_2=u,d$, the distinct B-O potentials
are specified by the isospin quantum number $I=0,1$.
The value of $\Lambda = |\lambda|$ is traditionally 
specified by an upper-case Greek letter: 
$\Sigma,\Pi,\Delta, \ldots$ for $\Lambda = 0,1,2,\ldots$.   
The value $+1$ or $-1$ of $\eta$ is traditionally specified 
by a subscript $g$ or $u$ on the upper-case Greek letter.  
In the case $\lambda = 0$, the value $+1$ or $-1$ of $\epsilon$ 
is traditionally specified 
by a superscript $+$ or $-$ on $\Sigma$.
Thus the flavor-singlet B-O potentials $V_\Gamma(r)$ are  labelled 
by $\Gamma = \Sigma_\eta^+,\Sigma_\eta^-,\Pi_\eta, \Delta_\eta, \ldots$, 
where $\eta$ is $g$ or $u$.  The light-field configurations with those energies
that are simultaneous eigenstates of 
$|\hat{\bm{r}}\cdot \bm{J}_{\rm light}|$, $(CP)_{\rm light}$, and  $R_{\rm light}$
can be labelled by $\Lambda_\eta^\epsilon$. 
We will refer to them as {\it Born-Oppenheimer configurations}.

In QCD without light quarks,
the ground-state flavor-singlet B-O potential $V_{\Sigma_g^+}(r)$ 
can be defined as the minimal energy of the gluon field 
in the presence of the static $Q$ and $ \bar Q$ sources.
An excited flavor-singlet B-O (or {\it hybrid}) potential $V_\Gamma(r)$,
can be defined as the minimal energy 
of the gluon field with quantum numbers $\Gamma$  
only if $V_\Gamma(r)$ does not exceed $V_{\Sigma_g^+}(r)$ 
by more than the mass of a glueball with the appropriate quantum numbers.  
In QCD with light quarks, the  minimal-energy prescription breaks down 
if $V_\Gamma(r)$ exceeds $V_{\Sigma_g^+}(r)$ 
by more than 2 or 3 times the mass of a pion,
depending on the quantum numbers $\Gamma$.  
It also breaks down if $V_\Gamma(r)$ exceeds twice the energy 
of a static meson, which is the minimal energy for
light-quark and gluon fields with the flavor of  a single  light quark
in the presence of a static $\bar Q$ source.
Similar complications arise in the definition of a flavor-nonsinglet  B-O
(or {\it tetraquark}) potential.
In all these cases, if the B-O potential  exists, it must be defined by
a more complicated prescription involving
excited states of the light fields with the specified quantum numbers.

Many of the hybrid potentials were calculated by Juge, Kuti, 
and Morningstar using quenched lattice QCD \cite{Juge:1999ie,Juge:2002br},
which does not include virtual quark-antiquark pairs.  
At large $r$, they approach linear functions of $r$.
At small $r$, they approach the repulsive Coulomb potential 
between a $Q$ and $\bar Q$ in a color-octet state,
which is approximately linear in $1/r$.
The deepest hybrid potentials are $\Pi_u$ and then $\Sigma_u^-$,
which is equal to the $\Pi_u$ potential at $r=0$.
In the limit $r \to 0$, the $Q$ and $\bar Q$ sources reduce to a 
local color-octet $Q \bar Q$ source, and
the B-O configuration reduces to a {\it static hybrid meson} or {\it gluelump},
which is a flavor-singlet state of the light fields bound to a static color-octet source. 
The gluelump spectrum was first calculated 
using quenched lattice QCD by Campbell, 
Jorysz, and Michael \cite{Campbell:1985kp}. 
It was recently calculated by Marsh and Lewis 
using lattice QCD with dynamical light quarks \cite{Marsh:2013xsa}. 
The ground-state gluelump
has $J^{PC}_{\rm light}$ quantum numbers $1^{+-}$. 
In the limit $r \to 0$, the $\Pi_u$ and $\Sigma_u^-$ potentials
differ from the repulsive Coulomb potential for a color-octet $Q \bar Q$ pair
by an additive constant  that can be interpreted as the energy 
of the ground-state $1^{+-}$ gluelump.

The tetraquark potentials can be specified by $\Lambda_\eta^\epsilon$
and quark flavors $q_1 \bar q_2$.
None of them have yet been calculated using lattice QCD. 
Some information about these potentials at small $r$
can be inferred from lattice QCD calculations of {\it static adjoint mesons},
which are flavor-nonsinglet states of the light fields
bound to a static color-octet source. 
Foster and Michael have calculated the adjoint meson spectrum 
using quenched lattice QCD 
with a light valence quark and antiquark  \cite{Foster:1998wu}.  
The $q \bar q$ adjoint mesons with the lowest energies are 
a vector ($J^{PC}_{\rm light}= 1^{--}$)
and a pseudoscalar ($0^{-+}$).  
The central values of their energies were larger than
that of the ground-state $1^{+-}$ gluelump by about 50~MeV
and 100~MeV, respectively,
but the differences were within the statistical errors.
Lattice QCD calculations with dynamical light quarks would be required to
determine definitively the ordering of the three energies.
For each adjoint meson, there must be tetraquark potentials 
that in the limit $r \to 0$ approach the repulsive Coulomb potential 
for  a color-octet $Q \bar Q$ pair
plus an additive constant that can be interpreted as the energy of the adjoint meson.
If these potentials remain well-defined at large $r$,
it is possible that they increase linearly with $r$,
like the flavor-singlet B-O potentials. In this case, 
the tetraquark potentials would have the same qualitative behavior 
as the hybrid potentials.

Given the quantum numbers $J^{PC}_{\rm light}$ of a $q \bar q$ adjoint meson,
we can deduce the corresponding B-O potentials.
The component $\hat{\bm{r}}\cdot \bm{J}_{\rm light}$ for an  adjoint meson with spin
$J_{\rm light}$ has $2J_{\rm light} + 1$ integer values ranging
from $-J_{\rm light}$ to $+J_{\rm light}$.  There must therefore be a
B-O potential for each integer value of $\Lambda$ from 0 up to $J_{\rm light}$.
The quantum number $\eta$ for the B-O potentials is the value of 
$(CP)_{\rm light}$ for the adjoint meson.
One of the B-O potentials is a $\Sigma$ potential with
reflection quantum number $\epsilon =  (-1)^{J_{\rm light}} P_{\rm light}$.
Thus the B-O potentials associated with the $1^{--}$ adjoint meson are 
$\Pi_g$ and $\Sigma_g^+$, while
the B-O potential associated with the 
$0^{-+}$ adjoint meson is $\Sigma_u^-$.
Since the $1^{--}$ and the $0^{-+}$ adjoint mesons have the lowest energies,
it is reasonable to expect the deepest tetraquark potentials 
to be $\Pi_g$, $\Sigma_g^+$, and $\Sigma_u^-$.  

There are several angular momenta that contribute to the spin vector $\bm{J}$ 
of a $Q \bar Q$ meson.  In addition to $\bm{J}_{\rm light}$,
there is the orbital angular momentum $\bm{L}_{Q \bar Q}$ 
and the total spin $\bm{S}$ of the $Q \bar Q$ pair.  
The spin vector of the meson can be expressed as
$\bm{J} = \bm{L} + \bm{S}$, where
$\bm{L} =  \bm{L}_{Q \bar Q} + \bm{J}_{\rm light}$.
The condition $\bm{r}  \cdot \bm{L}_{Q \bar Q} = 0$ implies
$\hat{\bm{r}} \cdot \bm{L} = \lambda$,
where $\lambda$ is the quantum number for $\hat{\bm{r}} \cdot \bm{J}_{\rm light}$.
This puts a lower limit on the quantum number $L$ for
$\bm{L}^2$: $L \ge \Lambda$. 
For a flavor-singlet $Q \bar Q$ meson with B-O configuration $\Lambda_\eta^\epsilon$, 
the parity and charge-conjugation quantum numbers are
\begin{subequations}
\begin{eqnarray}
P &=& \epsilon\,  (-1)^{\Lambda+L+1},
\label{Phybrid}
\\
C &=&  \eta\,  \epsilon\,  (-1)^{\Lambda+L+S}.
\label{Chybrid}
\end{eqnarray}	
\label{PChybrid}%
\end{subequations}
%

\begin{table}[t]
\begin{center}
\begin{tabular}{ccc|ccc}
\multicolumn{3}{c}{quarkonium hybrids} & \multicolumn{3}{c}{$Q \bar Qq \bar q$ tetraquarks} \\
~$\Gamma(nL)$~ & ~$S=0$~ & ~$S=1$~ & ~$\Gamma(nL)$~ & ~$S=0$~ & ~$S=1$~ \\
\hline
$\Pi_u^+(1P)$          & $1^{--}$         & $(0,{\bf 1},2)^{-+}$~        & 
~$\Pi_g^-(1P)$         & $1^{+-}$        & $(0,1,2)^{++}$                 \\
$\Pi_u^-(1P)$           & $1^{++}$       &  $({\bf 0},1,{\bf 2})^{+-}$~ & 
~$\Pi_g^+(1P)$        & ${\bf 1}^{-+}$ &   $({\bf 0},1,2)^{--}$          \\
$\Sigma_u^-(1S)$    & $0^{++}$        & $1^{+-}$                          & 
~$\Sigma_g^+(1S)$ & $0^{-+}$         & $1^{--}$                           \\
                                 &                       &                                        & 
~$\Sigma_u^-(1S)$  & $0^{++}$        & $1^{+-}$
\end{tabular}
\end{center}
\caption{Ground-state spin-symmetry multiplets for the two deepest hybrid potentials
$\Pi_u$ and $\Sigma_u^-$ and for the $\Pi_g$, $\Sigma_g^+$, 
and $\Sigma_u^-$ tetraquark potentials with light flavor $q \bar q$.
A bold $\bm{J}$ indicates that $\bm{J}^{PC}$
is an exotic quantum number that is not possible
if the constituents are only $Q \bar Q$.}
\label{tab:multiplets}
\end{table}

The $Q \bar Q$ mesons are conveniently organized into 
heavy-quark spin-symmetry multiplets consisting of states
with the same B-O configuration $\Lambda_\eta^\epsilon$,
radial quantum number $n$,
orbital-angular-momentum quantum number $L$, and flavors.
Each multiplet consists of a spin-singlet ($S=0$) state and 
either one or three spin-triplet ($S=1$) states.
Conventional quarkonia are energy levels in the 
flavor-singlet $\Sigma_g^+$ potential.
The spin-symmetry multiplet for the ground state in this potential consists of
a spin-singlet $0^{-+}$ state and a spin-triplet $1^{--}$ state.
The lowest-energy quarkonium hybrids are energy levels in the 
flavor-singlet $\Pi_u$ and $\Sigma_u^-$ potentials.
The ground-state spin-symmetry multiplets in these potentials are
given in Table~\ref{tab:multiplets}.
Tetraquark $Q \bar Q$ mesons are energy levels in B-O potentials 
labelled by $\Lambda_\eta^\epsilon$ and quark flavors.
The spin-symmetry multiplets for tetraquark $Q \bar Q$ mesons
are most easily specified by giving the 
$J^{PC}$ quantum numbers for tetraquark mesons with flavor $q \bar q$.
The ground-state spin-symmetry multiplets in the $\Pi_g$, $\Sigma_g^+$, 
and $\Sigma_u^-$ potentials are given in Table~\ref{tab:multiplets}.
The $J^{PC}$ quantum numbers 
are those for $I=0$ and $s \bar s$ tetraquarks
and for the neutral member of the $I=1$ isospin triplet.
The charged members of the $I=1$ triplet have the same $J^{P}$ 
and $G$-parity $G = -C$. 

Most of the observed decay modes of the $XYZ$ mesons
are hadronic transitions to a quarkonium.  
Selection rules for the hadronic transitions
provide constraints on the quarkonium hybrids 
or tetraquarks that can be considered as candidates
for specific $XYZ$ mesons.  The {\it spin selection rule} $S = S'$, where $S$ and $S'$ 
are the total spin quantum numbers for the $Q \bar Q$ pair
before and after the transition, follows from the approximate heavy-quark spin symmetry,
which is a consequence of the large mass of the heavy quark.
The {\it Born-Oppenheimer  selection rules} also require 
the B-O approximation, in which the 
hadronic transition between $Q \bar Q$ mesons proceeds through a transition 
between B-O configurations with fixed separation vector $\bm{r}$
for the $Q$ and $\bar Q$ sources.  For simplicity,
we will deduce these selection rules for transitions between neutral $Q \bar Q$ mesons 
with quantum numbers $J^{PC}$ and $J^{'P'C'}$.  The corresponding selection rules 
involving charged tetraquark mesons can then be deduced from isospin symmetry.
We consider a transition via  
the emission of a single hadron $h$ with quantum numbers $J_h^{P_h C_h}$
and orbital-angular-momentum quantum  number $L_h$.  
The conservation of the component of $\bm{J}_{\rm light}$ 
along the $Q \bar Q$ axis can be expressed as
$\lambda = \lambda' + \hat{\bm{r}} \cdot (\bm{J}_h + \bm{L}_h )$,
where $\bm{J}_h$ and $\bm{L}_h$ are the spin  and
orbital-angular-momentum vectors of  $h$.  
This constraint implies the selection rule
\begin{equation}
|\lambda - \lambda'| \le J_h  + L_h.
\label{lambdaselect}
\end{equation}	
Conservation of $(CP)_{\rm light}$  implies the selection rule
\begin{equation}
\eta = \eta' \cdot C_h P_h (-1)^{L_h}.
\label{CPselect}
\end{equation}	
If $\lambda = \lambda'=0$, there is an additional constraint
from invariance under reflection through a plane containing the $Q \bar Q$ axis:
\begin{equation}
\epsilon = \epsilon' \cdot P_h (-1)^{L_h}
\qquad (\lambda = \lambda'=0).
\label{Pselect}
\end{equation}	

We proceed to apply the selection rules to the $c \bar c$ $XYZ$ mesons. 
The spin selection rule implies that $XYZ$ mesons 
with transitions to the spin-singlet $\Sigma_g^+(1P)$ charmonium state  $h_c$ 
must be spin-singlet states
while those with transitions to the spin-triplet $\Sigma_g^+(1S)$ charmonium state $J/\psi$ 
or the spin-triplet $\Sigma_g^+(1P)$ charmonium state $\chi_{c1}$ must be spin-triplet states.
This puts strong constraints on $XYZ$ mesons 
with quantum numbers $1^{--}$.
In the hybrid multiplets in Table~\ref{tab:multiplets},
the only $1^{--}$ state is the spin-singlet state in  $\Pi_u^+(1P)$.
In the tetraquark multiplets in Table~\ref{tab:multiplets},
the only $1^{--}$ states are spin-triplet states 
in $\Pi_g^+(1P)$ and $\Sigma_g^+(1S)$.
The $1^{--}$ meson $Y(4220)$ decays 
into $h_c\, \pi^+ \pi^-$ \cite{Yuan:2013ffw}, so it must be a spin singlet.
It can only be identified with the $1^{--}$ state in the $\Pi_u^+(1P)$ multiplet
of the charmonium hybrid.
The $1^{--}$ meson $Y(4260)$ decays into $J/\psi\, \pi^+\pi^-$ \cite{Aubert:2005rm},
so it must be a spin triplet.
It can be identified with the $1^{--}$ state in either the $\Pi_g^+(1P)$ 
or $\Sigma_g^+(1S)$ multiplet of the isospin-0 charmonium tetraquark.

Several of the hadronic transitions for the neutral $c \bar c$ $XYZ$ mesons
are the emission of a single vector meson 
$\omega$ or $\phi$ with   $J_h^{P_h C_h} = 1^{--}$.
Since the kinetic energy of the vector meson is small compared to its mass,
we assume it is emitted in an $S$-wave state.
The B-O selection rules in Eqs.~(\ref{lambdaselect})--(\ref{Pselect}) reduce to
$|\lambda - \lambda'| \le 1$, $\eta = \eta'$, and 
also $\epsilon = - \epsilon'$ if $\lambda = \lambda'=0$.
If the final-state configuration is $\Sigma_g^+$ corresponding to quarkonium,
the selection rules reduce further to $\Lambda \le 1$, $\eta = +1$, and 
also $\epsilon =  -1$ if $\Lambda=0$.  The only 
possible initial-state configurations are $\Pi_g^-$, $\Pi_g^+$, and $\Sigma_g^-$.
This excludes all the hybrid multiplets in Table~\ref{tab:multiplets}.
The only tetraquark multiplets in Table~\ref{tab:multiplets}
that are allowed are $\Pi_g^-(1P)$ and $\Pi_g^+(1P)$.
The $X(3915)$ decays into $J/\psi\, \omega$
and has quantum numbers  $0^{++}$ \cite{Lees:2012xs}.
It can be identified with the $0^{++}$ state
in the $\Pi_g^-(1P)$ multiplet of isospin-0 charmonium tetraquarks.
The $Y(4140)$ decays into $J/\psi\, \phi$ \cite{Aaltonen:2009tz}
and therefore has $C=+$.  The $\phi$ in the final state 
suggests that the meson is an $s \bar s$ charmonium tetraquark.
The only spin-triplet $C=+$ states in Table~\ref{tab:multiplets} are the 
$0^{++}$, $1^{++}$, and $2^{++}$ states in $\Pi_g^-(1P)$.
The mass difference of about 230~MeV between the  $Y(4140)$
and the $X(3915)$ is approximately twice the difference 
between the constituent masses of the $s$ and lighter quarks.
This is compatible with the identifications of $Y(4140)$ and $X(3915)$ 
as states in $s \bar s$ and isospin-0 $\Pi_g^-(1P)$ multiplets, respectively.

The only hadronic transitions that have been observed for charged $XYZ$ mesons  
are the emission of a single pion with $J_h^{P_h C_h} = 0^{-+}$.
The Goldstone nature of the pion requires that it be emitted in a $P$-wave state.
The B-O selection rules in Eqs.~(\ref{lambdaselect})--(\ref{Pselect}) reduce to
$|\lambda - \lambda'| \le 1$, $\eta = \eta'$, and 
also $\epsilon = \epsilon'$ if $\lambda = \lambda'=0$.
If the final-state configuration is $\Sigma_g^+$ corresponding to quarkonium,
the selection rules reduce further to $\Lambda \le 1$, $\eta = +1$, and 
also $\epsilon =  +1$ if $\Lambda=0$.  The only 
possible initial-state configurations are $\Pi_g^-$, $\Pi_g^+$, and $\Sigma_g^+$.
Isospin symmetry requires the initial configuration to have isospin 1.  
The only tetraquark multiplets in Table~\ref{tab:multiplets} 
that are allowed are $\Pi_g^-(1P)$, $\Pi_g^+(1P)$, and $\Sigma_g^+(1S)$.
The $Z_c^+(3900)$ decays into $J/\psi\, \pi^+$  \cite{Ablikim:2013mio}.
Its neutral isospin partner $Z_c^0(3900)$ has $C=-$.
The spin-triplet $C=-$ tetraquark states in Table~\ref{tab:multiplets} are 
the $(\bm{0},1,2)^{--}$ states in $\Pi_g^+(1P)$ and the $1^{--}$ state in $\Sigma_g^+(1S)$.
The $Z_c^+(4020)$ decays into $h_c\, \pi^+$ \cite{Ablikim:2013wzq}.
Its neutral isospin partner $Z_c^0(4020)$ has $C=-$.
The only spin-singlet $C=-$ tetraquark state in Table~\ref{tab:multiplets} is the 
$1^{+-}$ state in $\Pi_g^-(1P)$.
The $Z_1^+(4050)$ decays into $\chi_{c1}\, \pi^+$  \cite{Mizuk:2008me}.  
Its neutral isospin partner $Z_1^0(4050)$ has $C=+$.
The spin-triplet $C=+$ tetraquark states in Table~\ref{tab:multiplets}
are the $0^{++}$, $1^{++}$, and $2^{++}$ states in $\Pi_g^-(1P)$.
The small mass difference between $Z_c(4050)$ and $Z_1(4020)$ 
is compatible with them being in the same $\Pi_g^-(1P)$ multiplet.

We used the spin selection rule to identify the $Y(4260)$ 
as a spin-triplet $1^{--}$ state
in either the $\Pi_g^+(1P)$ or  $\Sigma_g^+(1S)$ multiplet of isopin-0 charmonium tetraquarks.
We used the spin and B-O selection rules to identify
the $Z_c(3900)$ as a spin-triplet state
in either $\Pi_g^+(1P)$ or $\Sigma_g^+(1S)$ multiplet of isopin-1 charmonium tetraquarks.
The $\Pi_u$ hybrid potential is deeper than the $\Sigma_u^-$ hybrid potential.
If the $\Pi_g$ tetraquark potential is similarly deeper than the $\Sigma_g^+$ tetraquark potential, 
the most plausible identifications would be
$Z_c(3900)$ as a $\Pi_g^+(1P)$ state
and $Y(4260)$ as a $\Sigma_g^+(1S)$ state.

Our selection rules for hadronic transitions do not provide very useful constraints
on the few $b \bar b$ $XYZ$ mesons that have been observed. 
The $Z_b^+(10610)$ and $Z_b^+(10650)$ have transitions by emission of a single pion
into both the spin-triplet bottomonium states $\Upsilon(nS)$ 
and the spin-singlet bottomonium states  $h_b(nS)$   \cite{Belle:2011aa}.  
This violation of the spin selection rule
can be explained by the $Z_b^+(10610)$ having a large $B^* \bar B$ 
molecular component and the $Z_b^+(10650)$ having a large $B^* \bar B^*$ 
molecular component \cite{Bondar:2011ev,Cleven:2011gp,Mehen:2011yh}.
Within the B-O approach, the large molecular components 
would arise from energy levels in B-O potentials that are fortuitously
close to the $B^* \bar B$ and $B^* \bar B^*$ thresholds.

We have used the Born-Oppenheimer approximation to derive  selection rules 
for hadronic transitions between $Q \bar Q$ mesons.  They strongly constrain
the $c \bar c$ $XYZ$ mesons that can be candidates for ground-state energy levels
in the B-O potentials for charmonium hybrids and tetraquarks.
The selection rules should provide valuable guidance in the search for additional $XYZ$ states 
through their hadronic transitions.
Lattice QCD calculations of the tetraquark potentials are needed to confirm 
that the deepest potentials have been correctly identified.
They would also allow the B-O approximation to be developed
into a quantitative theoretical framework
for understanding the  $XYZ$ mesons that is  based firmly on QCD.

\begin{acknowledgments}
This research was supported in part by the Department of Energy 
under grant DE-FG02-05ER15715 and  by the
National Science Foundation under grant PHY-1310862.
\end{acknowledgments}


%

\end{document}